\documentclass[acmsmall,screen]{acmart}

\usepackage{enumitem}
\usepackage{xcolor}
\usepackage{tabularx}
\usepackage{booktabs}
\usepackage{array}
\usepackage{calc}
\usepackage{multirow}
\usepackage{wrapfig}
\usepackage{arydshln}  

\begin{document}

\title[Automated Soap Opera Testing Directed by LLMs and Scenario Knowledge: Feasibility, Challenges, and Road Ahead]{Automated Soap Opera Testing Directed by LLMs and Scenario Knowledge: Feasibility, Challenges, and Road Ahead}

\author{Yanqi Su}
\email{Yanqi.Su@anu.edu.au}
\affiliation{%
  \institution{The Australian National University}
  \country{Australia}
}

\author{Zhenchang Xing}
\email{Zhenchang.Xing@data61.csiro.au}
\affiliation{%
  \institution{CSIRO's Data61}
  \country{Australia}
}

\author{Chong Wang}
\email{chong.wang@ntu.edu.sg}
\affiliation{%
  \institution{Nanyang Technological University}
  \country{Singapore}
}

\author{Chunyang Chen}
\email{chun-yang.chen@tum.de}
\affiliation{%
  \institution{Technical University of Munich}
  \country{Germany}
}

\author{Xiwei Xu}
\email{xiwei.xu@data61.csiro.au}
\affiliation{%
  \institution{CSIRO's Data61}
  \country{Australia}
}

\author{Qinghua Lu}
\email{qinghua.lu@data61.csiro.au}
\affiliation{%
  \institution{CSIRO's Data61}
  \country{Australia}
}

\author{Liming Zhu}
\email{liming.zhu@data61.csiro.au}
\affiliation{%
  \institution{CSIRO's Data61 \& 	University of New South Wales}
  \country{Australia}
}









\begin{abstract}
  Exploratory testing (ET) harnesses tester's knowledge, creativity, and experience to create varying tests that uncover unexpected bugs 
from the end-user's perspective.
Although ET has proven effective in system-level testing of interactive systems, the need for manual execution has hindered large-scale adoption.
In this work, we explore the feasibility, challenges and road ahead of automated scenario-based ET (a.k.a soap opera testing).
We conduct a formative study,
identifying key insights for effective manual soap opera testing and challenges in automating the process. 
We then develop a multi-agent system leveraging LLMs and a Scenario Knowledge Graph (SKG) to automate soap opera testing. 
The system consists of three multi-modal agents, Planner, Player, and Detector that collaborate to execute tests and identify potential bugs.
Experimental results demonstrate the potential of automated soap opera testing,
but there remains a significant gap compared to manual execution, especially under-explored scenario boundaries and incorrectly identified bugs.
Based on the observation, we envision road ahead for the future of automated soap opera testing, focusing on three key aspects: the synergy of neural and symbolic approaches, human-AI co-learning, and the integration of soap opera testing with broader software engineering practices. 
These insights aim to guide and inspire the future research.


\end{abstract}





\maketitle

\section{Introduction}Exploratory testing (ET)~\cite{exploratory_tutorial,exploratory_nature,exploratory_book_tips_tricks_tours_techniques_to_guide_test_design} leverages the knowledge, creativity, and experience of testers to create varying and diverse tests on the fly to provoke and identify unexpected bugs.
Unlike scripted testing, where test cases are executed 
in a mechanical and repetitive way, ET allows for open-ended results~\cite{exploratory_tutorial,exploratory_nature,exploratory_book_tips_tricks_tours_techniques_to_guide_test_design}. 
Some outcomes may be expected and predicted; others may not. 
Testers engage by configuring, operating, observing, and evaluating the system, critically examining outcomes, and reporting any potential bugs.
Although scripted testing has received much attention in the research community, 
ET is widely practiced and highly valued in the software industry~\cite{exploratory_comparison_test_case,exploratory_a_multiple_case_study,exploratory_how_exploratory_testing_used,exploratory_tester_knowledge}.
Some scientific studies~\cite{exploratory_a_multiple_case_study,exploratory_tester_knowledge,exploratory_comparison_test_case} show ET is an effective and efficient testing approach for system-level testing of interactive systems through the GUI and from the end user's viewpoint, and can find functional, usability, performance and security bugs~\cite{exploratory_how_exploratory_testing_used,syskg}. 

Cem Kaner and others~\cite{exploratory_tutorial,exploratory_book_tips_tricks_tours_techniques_to_guide_test_design,exploratory_nature} have established concrete principles for ET, emphasizing that not scripting does not mean not preparing~\cite{exploratory_tutorial}. 
Exploratory testers need to gather, analyze, and study extensive information about the application domain, software products, and users.
Scenario-based ET, also known as soap opera testing~\cite{soap_opera_testing}, involves designing complex test scenarios to provoke failures and recognize them.
Recent work~\cite{syskg, soapoperatg, su2024enhancing} has constructed a system knowledge graph (SYSKG) from bug reports, modeling user tasks, failures, and system interactions
to support soap opera testing. 
These studies generate soap opera tests by combining relevant scenarios from SYSKG. 
Experimental results~\cite{syskg, soapoperatg, su2024enhancing} have shown that these tests help testers learn the product, simulate expert users, and uncover hidden failures.
However, the need for manual execution has limited the large-scale adoption of soap opera testing. 
Leveraging the capabilities of large language models (LLMs) in natural language and vision understanding, this work explores the feasibility, challenges, and road ahead of automated soap opera testing guided by LLMs and scenario knowledge.

We conduct a formative study to investigate how humans effectively perform soap opera testing and to identify the challenges in automating this process. 
Six professionals specializing in soap opera testing participated in the study.
First, a user study tasks each participant with conducting soap opera testing using provided soap opera tests, aiming to uncover as many bugs as possible. 
Following this, we conduct interviews that reference both the bug discovery process and the participants' prior experience to identify the insights and challenges.
Two insights emerge for effective soap opera testing: (1) \textit{discovering unexpected behaviors while exploring and observing}, and (2) \textit{expanding exploration boundaries through creative thinking}. 
This means that testers need to remain observant at all stages of the soap opera test execution, continuously identifying any unexpected behaviors to uncover potential bugs.
In addition, while ensuring the soap opera test is completed, testers need to engage in creative thinking to explore how different inputs affect the system and whether various side tasks along the soap opera test might trigger bugs. 
It's similar to a role-playing game where, alongside completing the main quest, players can unlock side quests.



The challenges of automated soap opera testing are soap opera test auto-play and real-time bug auto-detection. 
Soap opera tests, written in natural language (NL), often suffer from quality issues such as missing UI instructions and outdated UI descriptions, complicating the automation process. 
Additionally, effective bug detection requires the system to understand both the executed instructions and the resulting UI changes in real-time, necessitating multi-modal understanding.
Addressing these challenges is crucial for the effectiveness of automated soap opera testing.
To explore feasibility, we developed a multi-agent system inspired by two insights, leveraging the advanced understanding capabilities of multi-modal LLMs in both NL and vision, combined with scenario knowledge to address identified challenges.
Specifically, 
this system consists of three multi-turn-dialogue agents, namely Planner, Player, and Detector, powered by multi-modal LLMs and a scenario knowledge graph (SKG). 
Given a soap opera test (a list of system operation steps) and the initial GUI status, three agents collaborate to perform the testing for identifying any potential bugs.
To evaluate the feasibility of automated soap opera testing, we tested 3 apps, each with 10 soap opera tests, resulting in the submission of 34 bug reports. 
Of these, 3 bugs were fixed, 13 confirmed, 6 assigned priority and severity, and 1 marked as duplicate. 
These results demonstrate the potential of automated soap opera testing,
but there remains a significant gap compared to manual execution, such as under-explored scenario boundaries and incorrectly identified bugs (false positives). 
To better understand the high false positive rate, we analyzed the incorrectly identified bugs and identified four main causes: LLM hallucinations, screenshot-based GUI presentation, incorrect GUI element location, and unreasonable or unnecessary enhancement suggestions.



By the preliminary exploration of automated soap opera testing, we summarize key observations and outline the road ahead for future research.
First, the synergy of neural and symbolic approaches is essential for effective soap opera testing. 
By integrating app-specific knowledge at various aspects, LLMs can enhance the understanding of software under test, thereby broadening and deepening exploration.
Second, while humans provide valuable insights for AI in automated soap opera testing, AI’s behavior can also inspire human creativity, uncovering bugs. 
As we transition from manual to automated testing, a Human-AI collaborative approach could serve as an effective middle ground.
Lastly, rather than stopping after discovering a single bug, soap opera testing excels at exploring further, uncovering related issues and identifying different manifestations of the same root cause. 
Therefore, combining soap opera testing with broader software engineering knowledge, such as code analysis for root cause identification, is a logical next step.
Our contributions are:
\begin{itemize}[leftmargin=*]
	\item We conduct a formative study to identify the insights of effective manual soap opera testing and the challenges of automated soap opera testing.
	
	\item We design a multi-agent system based on LLMs and scenario knowledge for the preliminary feasibility exploration of automated soap opera testing.
	
	\item Through extensive experimental analysis, we find that automated soap opera testing directed by LLMs and scenario knowledge is promising, but there is still a significant gap compared to manual soap opera testing.
	
	\item We outline the road ahead for automated soap opera testing, offering guidance for future research. 
\end{itemize}

\section{Formative Study}


We conduct a formative study to investigate how humans effectively perform soap opera testing and identify the challenges in automating this process, as follows:




\begin{description}

        \item[RQ1:] How effective is manual soap opera testing?
        
        
        \item[RQ2:] How can manual soap opera testing be conducted effectively? 
	  \item[RQ3:] What are the challenges in automating soap opera testing?
\end{description}

\subsection{Study Setup}
We recruited six professionals specializing in soap opera testing as participants, with four from industry and two from academia. 
Each participant has a minimum of 2 years of experience in soap opera testing for large interactive systems with GUI, such as desktop or mobile apps.
Notably, four participants have experience in GUI-based automated Android testing.
First, a user study tasks each participant with conducting soap opera testing to uncover bugs. 
Then, we conduct interviews based on the bug discovery process and the participants' prior experience, focusing on: (a) insights for effectively doing soap opera testing, and (b) challenges in automating this process.   

\subsubsection{User Study}
Each participant is tasked with conducting soap opera testing using five predefined soap opera tests, aiming to uncover as many bugs as possible within 3 hours. 

\textbf{Dataset.} 
Three open-source Android apps were selected as our software under test (SUT) from different categories:
\href{https://play.google.com/store/search?q=firefox&c=apps&hl=en}{Firefox}, a fast and privacy-focused browser;
\href{https://play.google.com/store/search?q=wordpress&c=apps&hl=en}{WordPress}, a website builder for creating posts and tracking analytics; and
\href{https://play.google.com/store/search?q=antennapod&c=apps&hl=en}{AntennaPod}, a podcast manager and player.
These apps were selected based on the 8 candidate apps from a recent study~\cite{dataset_functionalbugs} on Android functional bugs, taking into account the scale of issues and active developer engagement. All apps include over 1 million downloads on Google Play (popularity), a public issue tracking system (traceability), and approximately ten years of development history (trustworthiness).
For each app, we construct a dataset of 10 soap opera tests covering diverse features, such as login, bookmarks, privacy, browser engine, and translation. 
Each test consists of a series of system operation steps sourced from various origins: user manuals, issues, pull requests and bug reports.
The average number of steps per test is 5.9 for Firefox, 5.0 for WordPress, and 4.7 for AntennaPod.
To ensure continuity in testing, each participant is assigned 5 soap opera tests from a app. 
Two participants, who had previously used Firefox, are assigned this app. 
The remaining four participants, who had no prior experience with the apps, are randomly assigned to the other apps. 
This arrangement helps unveil the insights and challenges of soap opera testing when the testers is familiar or unfamiliar with the SUT.








\subsubsection{Interviews}
We conduct interviews with participants based on the bug discovery process from user study and their prior experience. 
This provides a fresh and concrete foundation for interviews, which aims at addressing the two research questions.
Then, we use thematic analysis~\cite{braun2006_thematic_analysis} to summarize and synthesize the content of the interviews, leading to the final conclusions.

First, \textit{How can manual soap opera testing be conducted effectively?} 
Participants are asked to recount their bug discovery process in the user study, including how these bugs were triggered and identified through soap opera testing. Participants may back up their claims with prior testing experience.
Second, \textit{What are the challenges in automating soap opera testing?}
Participants share their insights, referring to specific cases in the user study and also drawing on their general experiences with both soap opera testing and GUI-based automated testing where applicable. 
The discussion centers on the challenges associated with triggering and identifying bugs within the context of automation.

\subsection{RQ1: Effectiveness of Manual Soap Opera Testing}
The six participants found 11, 16, 5, 9, 8, and 8 bugs across Firefox, WordPress, and AntennaPod, respectively.
The status of identified bugs is 5 bugs fixed, 28 bugs confirmed, 4 bugs assigned priority and severity, 5 bugs as wontfix.
These bugs span a diverse range of categories, including functional, usability, performance, crash issues.
We also received numerous acknowledgments from the development teams, such as: ``Thank you especially for submitting so many detailed issues'' and some bugs (e.g., \href{https://github.com/AntennaPod/AntennaPod/issues/7357}{Issue7357}) also sparked discussions among developers.

\subsection{RQ2: Insights from Manual Soap Opera Testing}
Through the formative study, we identify two key insights for effectively conducting soap opera testing: discovering unexpected behaviors while exploring and observing, and expanding the boundaries of exploration by creative thinking.
These two insights echo the key findings in~\cite{exploratory_book_tips_tricks_tours_techniques_to_guide_test_design}: \textit{An exploratory tester who uses a scenario as a guide will often pursue interesting alternative inputs or pursue some potential side effect that is not included in the script. -- James A. WHITTAKER.}


\subsubsection{Discover unexpected behaviors while exploring and observing.}\label{subsubsec:Discover unexpected behaviors while exploring and observing.}
All six participants mention that unlike scripted testing, where bugs are detected based on predefined oracles after executing test scripts, soap opera testing enables the detection of unexpected bugs during the execution of intermediate steps, even beyond the known oracles.
An example from Participant 2 ($P2_{Firefox}$) for Firefox on \href{https://support.mozilla.org/en-US/kb/how-use-find-page-firefox-android}{Find in Page function} from the user manual illustrates this practice well. 
When executing the middle step (\textit{Enter your search term ...}
) of this soap opera test, $P2_{Firefox}$ discovered that the ``Find in Page'' feature incorrectly starts from `4/15' instead of `1/15' when the page is at the top, as reported in \href{https://bugzilla.mozilla.org/show_bug.cgi?id=1913291}{Bug1913291}.
$P2_{Firefox}$ then proceeded to the next step (\textit{Navigate through search results using the up or down ...}). 
$P2_{Firefox}$ found that ``Find in page'' inconsistently fails to highlight certain search results, leading to confusion and difficulty in locating terms (\href{https://bugzilla.mozilla.org/show_bug.cgi?id=1913295}{Bug1913295}). 
Based on this confusing behavior, $P2_{Firefox}$ scrolled the screen to locate the highlighted search results, discovering that scrolling while searching triggers the selection tool panel in ``Find in Page'' (\href{https://bugzilla.mozilla.org/show_bug.cgi?id=1913304}{Bug1913304}). 
Additionally, navigating through search results caused a blue box to appear around buttons or links (\href{https://bugzilla.mozilla.org/show_bug.cgi?id=1913307}{Bug1913307}).
Furthermore, tapping on the highlighted search result or a blank area caused the highlight to disappear
(\href{https://bugzilla.mozilla.org/show_bug.cgi?id=1913299}{Bug1913299}).
An error displaying ``0/0'' occurred when rapidly tapping the ``Up/Down'' buttons, but this issue had already been reported by others. 
However, since the steps to reproduce this bug have changed, we submitted a comment (\href{https://bugzilla.mozilla.org/show_bug.cgi?id=1807147#c11}{Bug1807147\#Comment11}) to update
based on the new observations made by $P2_{Firefox}$.
After reporting these bugs, 4 have already been confirmed. For details, please refer to the bug links mentioned above.

This example also reveals another significant issue. 
\href{https://bugzilla.mozilla.org/show_bug.cgi?id=1807147}{Bug 1807147} was reported two years ago, but related issues in the ``Find in Page'' feature were missed. Developers patched only the reported bug. 
Had exploratory testing (ET) been conducted thoroughly, other issues could have been identified, leading to a more systematic resolution. 
This encourages deeper investigation of root causes, shifting from isolated bug fixes to comprehensive problem-solving.

\subsubsection{Expand the boundaries of exploration by creative thinking.}\label{subsubsec:insights2_expand_boundaries}

In addition to staying observant for any unexpected behaviors during the soap opera tests, participants also use their creativity and experience to engage in creative thinking. 
This allows them to generate various inputs and branches to expand the boundaries of exploration in soap opera testing. 
This is similar to exploring side quests in a role-play game while the player completes the main mission~\cite{syskg}.

\textbf{Input Exploration.}
$P1_{Firefox}$ and $P2_{Firefox}$ emphasize the importance of diverse input exploration in triggering bugs during soap opera testing, which
should be guided by the current scenario, leveraging experience and known bug-finding techniques to create representative inputs.
For instance, $P2_{Firefox}$ encountered a scenario requiring a website input while executing a soap opera test. 
Instead of entering the correct website, $P2_{Firefox}$ intentionally misspelled it, which triggered the issue ``Entering a misspelled website causes the loading to stall, and refreshing reverts to the previously visited page'' (\href{https://bugzilla.mozilla.org/show_bug.cgi?id=1913318}{Bug1913318}).
In another example, when the soap opera test involved adding a search engine, $P2_{Firefox}$ attempted to add a search engine with the same name and details as an existing one. 
This led to the discovery of a bug, ``Adding a search engine allows duplicate entries''
(\href{https://bugzilla.mozilla.org/show_bug.cgi?id=1913414}{Bug1913414}).
After reporting bugs, one was assigned severity and one was confirmed.
When $P1_{Firefox}$ needed to edit a saved login, $P1_{Firefox}$ utilized the boundary value analysis technique by entering an extremely long username and password. 
This triggered the bug ``Long usernames and passwords overlap with the `Clear' button...'' (\href{https://bugzilla.mozilla.org/show_bug.cgi?id=1912207}{Bug1912207}), assigned severity.


\textbf{Branch Exploration.}
All participants noted that while ensuring the soap opera test is executed, performing random or logically driven side tasks can be beneficial for triggering bugs.

\textit{ 
Random Branch Exploration Triggering bugs.}
After $P5_{AntennaPod}$
completed the soap opera test and discovered a bug (\href{https://github.com/AntennaPod/AntennaPod/issues/7349}{Issue7349}: ``After removing a podcast, its episode stays in the miniplayer''), $P5_{AntennaPod}$ randomly clicked a button on the current page, triggering a crash (\href{https://github.com/AntennaPod/AntennaPod/issues/7350}{Issue7350}).
Similarly, when $P6_{AntennaPod}$ was executing a soap opera test following the steps in \href{https://github.com/AntennaPod/AntennaPod/issues/5822}{Issue5822}, after performing ``Remove Podcast'', a bug was triggered where ``The removed podcast still appears under `Get surprised' after 
confirming removal'',
reported in \href{https://github.com/AntennaPod/AntennaPod/issues/7370}{Issue7370} and subsequently fixed.
After discovering this bug, $P6_{AntennaPod}$ became curious about what would happen after clicking on the removed podcast under ``Get surprised'', which triggered another crash
(\href{https://github.com/AntennaPod/AntennaPod/issues/7371}{Issue7371}).

\textit{Logically Driven Side Tasks Triggering Bugs.}
For a soap opera test on translation, after clicking `Translate' to translate the entire page, $P2_{Firefox}$ noticed that the translation process 
toke
unusually long. 
To cancel the process, $P2_{Firefox}$ spontaneously clicked the `Not now' button, but
fails to cancel translation (\href{https://bugzilla.mozilla.org/show_bug.cgi?id=1913602}{Bug1913602}).
$P2_{Firefox}$ then observed that ``The page translation occurs in stages, with an initial partial translation followed by a delayed translation of the remaining content'' (\href{https://bugzilla.mozilla.org/show_bug.cgi?id=1913606}{Bug1913606}).
Based on this observation, $P2_{Firefox}$ attempted to refresh the page to complete the partial translation but discovered another issue: ``Refreshing a translated webpage reverts it to the original language...'' (\href{https://bugzilla.mozilla.org/show_bug.cgi?id=1915093}{Bug1915093}).
After exploring these branches, $P2_{Firefox}$ resumed the main task of soap opera test by tapping the `Reader view' icon to return the webpage to normal mode. 
This revealed
``Switching between Reader view and normal mode causes the page to revert to the original language...'' (\href{https://bugzilla.mozilla.org/show_bug.cgi?id=1913604}{Bug1913604}).
$P2_{Firefox}$ also noticed that ``The `Translation' icon color on the address bar changes inconsistently'' (\href{https://bugzilla.mozilla.org/show_bug.cgi?id=1913605}{Bug1913605}).
Four of these bugs have already been confirmed.

$P1_{Firefox}$ used the equivalence class to trigger bugs. 
After discovering \href{https://bugzilla.mozilla.org/show_bug.cgi?id=1912207}{Bug1912207} (``Long usernames...overlap with the `Clear'...when editing saved logins''), $P1_{Firefox}$ realized that the `Add password' page also has username and password input fields, which might exhibit the same issue.
$P1_{Firefox}$ then navigated to the `Add password' page and attempted to input long usernames and passwords, which led to \href{https://bugzilla.mozilla.org/show_bug.cgi?id=1912199}{Bug1912199}: ``Input Field Overlap with `Clear' Button in `Add Password' ''.


\vspace{1mm}
\noindent\fbox{\begin{minipage}{13.7cm}
Two key insights were identified for effectively conducting soap opera testing: discovering unexpected behaviors while exploring and observing, and expanding the boundaries of exploration by creative thinking. 
Testers need to remain observant at all stages of the soap opera test execution, continuously identifying any unexpected behaviors.
Testers also utilize their creativity and experience to generate diverse inputs and explore different branches, triggering unexpected behaviors. 
These insights are further validated by~\cite{exploratory_tutorial,exploratory_book_tips_tricks_tours_techniques_to_guide_test_design}
and supported by ~\cite{su2024enhancing,syskg,soapoperatg}.
 \end{minipage}}\\
\vspace{1mm}

\subsection{RQ3: Challenges in Automated Soap Opera Testing}
Automated soap opera testing requires not only automatically executing the soap opera tests (Soap Opera Test Auto-play) but also identifying unexpected bugs that may arise at any point during execution (Real-Time Bug Auto-detection).
\subsubsection{Soap Opera Test Auto-play}\label{subsubsec:SoapOperaTestAuto-Play}
All participants identified a key challenge in mapping NL steps in soap opera tests to the corresponding UI instructions for automated manipulation.

\textbf{Missing UI Instructions.}
Missing UI instructions refers to situations where some necessary UI instructions are skipped between two steps, or where a single step in the soap opera test requires multiple UI instructions
to be executed.
Even user manuals can contain instances of missing UI instructions. 
For example, when $P2_{Firefox}$ executed a soap opera test from the user manual \href{https://support.mozilla.org/en-US/kb/add-delete-and-view-bookmarked-webpages-firefox-android#w_to-move-a-bookmark-to-a-new-folder}{To move a bookmark to a new folder}, the move operation failed despite following all the steps. 
Initially, $P2_{Firefox}$ assumed this was a bug, but after expanding the boundaries of exploration, $P2_{Firefox}$ discovered that the failure was due to a missing step in the user manual. 
Specifically, a step states, ``Tap the `Back' arrow ... a few times until you return to browsing the web''. 
During this process of tapping `Back' a few times, the crucial UI instruction (``Tap the `Save' icon to save the change'') is missing, which is necessary to complete the move operation successfully.
This is an example where automated execution could lead to a false positive, resulting in the reporting of a non-existent bug.
In a soap opera test from \href{https://bugzilla.mozilla.org/show_bug.cgi?id=1812113}{Bug1812113}, the step “Open recently closed tabs” left $P1_{Firefox}$ unsure how to proceed, even with prior Firefox experience. 
Searching \href{https://bugzilla.mozilla.org/home}{Bugzilla}, $P1_{Firefox}$ found detailed instructions in \href{https://bugzilla.mozilla.org/show_bug.cgi?id=1881509}{Bug1881509}: ``Open Tabs Tray'', ``Tap on the three dots menu'', and ``Select `Recently closed tabs'''. 
Without sufficient execution details, automated testing may halt at such steps.


\textbf{Outdated UI Description.} 
Outdated UI descriptions occur when soap opera test steps no longer match the app's current interface due to updates. 
For example, ``Logins and passwords'' is now ``Passwords'', and ``Saved logins'' is now ``Saved passwords''. 
Similarly, ``Navigate to History from Settings'' is now under the ``Three-dot'' menu. 
These discrepancies hinder automated execution.

\subsubsection{Real-Time Bug Auto-detection}\label{subsubsec:Real-TimeBugAuto-detection}
All participants agreed that effective automated real-time bug detection requires both image recognition and NL understanding. 
Specifically, automated soap opera testing needs to comprehend the UI instructions being executed and the corresponding changes triggered in the UI.
From the formative study, bugs were categorized into two types: inner-page, identified within a single UI (e.g., long usernames overlapping with the 'Clear' button), and inter-page, discovered by observing changes across multiple UI pages and actions that triggered these changes (e.g., \href{https://bugzilla.mozilla.org/show_bug.cgi?id=1913604}{Bug1913604}, where switching modes causes the page to lose translation).

\vspace{1mm}
\noindent\fbox{\begin{minipage}{13.7cm}
The success of automated soap opera testing depends on two key challenges: ensuring accurate test auto-play and achieving real-time bug detection. Variability in test steps, such as missing or outdated UI instructions, complicates automation and can cause failures. Effective bug auto-detection also requires real-time understanding of both the executed instructions and UI changes. 
 \end{minipage}}\\
\vspace{1mm}





\section{Preliminary Exploration of Automated Soap Opera Testing}\label{sec:approach}
To explore the feasibility of automated soap opera testing, we developed a multi-agent system inspired by two key insights uncovered in the formative study, 
leveraging the superior understanding capabilities of multi-modal LLMs in both NL and vision, combined with scenario knowledge, to address the identified challenges.

\subsection{Automated Multi-Agent Workflow.}
As shown in Figure~\ref{fig:approach_workflow}, 
the system consists of three \textit{multi-turn-dialogue} agents, namely \textit{Planner}, \textit{Player}, and \textit{Detector}, powered by multi-modal LLMs and a scenario knowledge graph (SKG). 
Given a soap opera test (a list of system operation steps) and the initial GUI status (a screenshot), the three agents collaborate to perform the testing as follows:

\begin{figure*}
    \centering
    \setlength{\abovecaptionskip}{-0cm}
    \setlength{\belowcaptionskip}{-0.4cm}
    \includegraphics[width=1
    \textwidth, height=4.6cm]{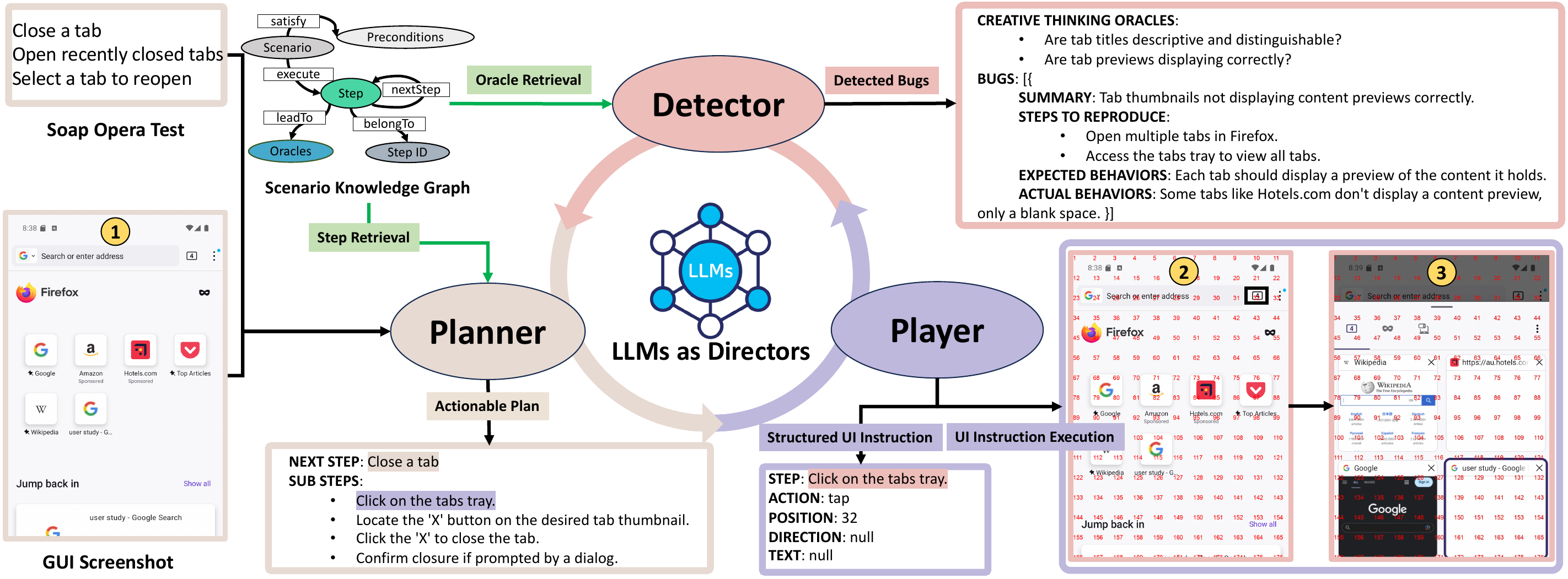}
    \begin{minipage}{1.0\textwidth} 
    \end{minipage} 
    \caption{Approach Overview of Automated Soap Opera Testing}
    \label{fig:approach_workflow}
    \vspace{-0.4cm}
\end{figure*}


\textbf{Planner:} 
Soap opera tests often face issues like missing steps or outdated UI descriptions, making them unsuitable for direct use. 
To address this, the Planner agent generate an actionable plan (detailed sub-steps) by utilizing LLMs based on the test description, the GUI status and step knowledge from SKG. 
As illustrated in Figure~\ref{fig:approach_workflow}, based on the given test and the current GUI status (Screenshot 1),
Planner identifies that the step ``Close a tab'' needs to be executed. 
Planner then generates a plan for it based on the retrieved steps in SKG, which includes four sub-steps, e.g., ``Click on the tabs tray'', ``Locate...'', which are more detailed and suitable for the current GUI status.
    
\textbf{Player:} From the generated plan, the Player agent processes one sub-step at a time,
translating it into a structured UI instruction 
based on the LLMs' understanding of both the sub-step and the current GUI status. 
Player then 
executes the instruction and inspects the resulting GUI status by taking a screenshot. 
For example, for the sub-step ``Click on the tabs tray'' and the current GUI status (Screenshot 2) in Figure~\ref{fig:approach_workflow}, Player translates the sub-step into a structured instruction with a ``tap'' action targeting element number ``32'' (indicating the position to tap). 
This instruction is then executed,
advancing the GUI to the next status (Screenshot 3).
    
\textbf{Detector:} 
The Detector agent is employed after each Player execution to identify potential unexpected bugs that may occur at any point.
Detector first retrieves relevant oracle knowledge from SKG based on the executed UI instruction. 
Then, using the GUI statuses before and after the Player's execution, along with the executed sub-step and retrieved oracles, LLMs further do creative thinking to generate more potential oracles.
Afterward, Detector assesses whether the subsequent GUI status violates any of these oracles, both retrieved and generated.
If oracle violations are detected, bugs are identified.
For example, Figure~\ref{fig:approach_workflow} illustrates a detected bug ``Tab thumbnails not displaying content previews correctly'' after executing the sub-step ``Click on the tabs tray''. 
After reporting in \href{https://bugzilla.mozilla.org/show_bug.cgi?id=1912739}{Bug1912739}, this bug was confirmed.
Notably, the retrieved oracles from the SKG included ``Thumbnails should be displayed properly ...'', providing support for bug detection.  

\textbf{Loop Iteration.} The above process iterates until Planner completes all steps in the given soap opera test, with Player executing instruction and Detector performing bug detection for each sub-step. 
To ensure the agents remain aware of the complete testing context, they are designed to reserve the
previous query-response records (e.g., for Planner, query is the test description and GUI status, response is the generated plan). These records are 
used to generate responses in subsequent turns. 
To satisfy the requirement of analyzing test description and GUI simultaneously, the LLMs employed in the three agents are \textit{multi-modal models}, such as \href{https://platform.openai.com/docs/models/gpt-4o}{GPT-4o}. 

\subsection{Application-Specific Knowledge Augmentation.}

While LLMs excel in understanding NL and vision, they lack application-specific knowledge, such as operation steps and oracles, needed for effective test auto-play and bug auto-detection. 
Our user study shows participants often rely on the software’s repository for UI operations and bug identification. 
To address this, we construct a scenario knowledge graph (SKG) with operation steps and oracles from crowd-sourced sources like bug reports and issues, using a retrieval-augmented generation (RAG)~\cite{rag_lewis2020retrieval} to provide this knowledge to Planner and Detector.

\subsubsection{Scenario Knowledge Graph}
In this work, SKG is constructed by adapting SYSKG pipeline~\cite{su2024enhancing}.
As shown in Figure~\ref{fig:approach_workflow}, SKG consists of scenarios, with each entry containing the following key information extracted from various sections of a bug report:
\textbf{Summary} as a high-level description of the scenario, extracted from the ``title'' section;
\textbf{Preconditions} as descriptions indicating the prerequisites required to execute steps, extracted from the ``prerequisites'' section;
\textbf{Steps} as normalized step-by-step descriptions of the scenario, extracted from the ``S2Rs'' section;
\textbf{Oracles} as descriptions of the expected behaviors, extracted from the ``EBs'' section.
The steps and oracles serve as the primary sources for the KB-based RAG in our Planner and Detector, while the summary and preconditions provide essential context to enable more accurate knowledge retrieval. 
Note that ``normalized'' for steps means they are deduplicated by clustering descriptions with the same meaning. 
Each normalized step is assigned a unique ID and includes a set of alternative descriptions, along with a corresponding set of associated oracles. (cf. the construction of SYSKG~\cite{su2024enhancing}). 
For example, a normalized step with ID `8' in Firefox has 47 alternative descriptions, including ``Go to Tabs Tray'' and ``Open tab tray'', and it has 19 corresponding oracles, such as ``Tab thumbnails not displaying content previews
correctly'' and ``Tooltip should display explanatory text''.
The SKG for Firefox is built from 7,947 \href{https://bugzilla.mozilla.org/buglist.cgi?query_format=advanced&resolution=---&classification=Client%20Software&product=Fenix&list_id=17225992}{Bugzilla} bug reports.
For WordPress, it uses 21,010 GitHub issues and pull requests.
AntennaPod’s SKG is based on 7,317 GitHub issues and pull requests.
\subsubsection{KG-Based Retrieval-Augmented Generation}\label{sec:rag}
To leverage the crowd-sourced knowledge about the SUT, this system integrates the constructed KG and employs 
RAG.
For efficient retrieval, we transform the SKG into vector representations using chunking and embedding process. 
As the SKG consists of nodes and relations, we convert it into structured text, where each node’s relations are represented as attributes.
For example, each step node is converted into structured text with two key properties: oracles and step ID. This structured text is then split into chunks for vectorization, with a fixed token chunk size and overlap to ensure context continuity and minimize information loss at chunk boundaries.
For embedding, we utilize an embedding model to convert each chunk into a vector representation, capturing the rich textual information from the steps and oracles, ensuring optimal retrieval performance. 
The vectorized SKG enables the agent to search the KG using both keyword-based and semantic queries.
Based on the retrieved results, the agent can generate outputs that are augmented with knowledge from the SKG. 
We currently use the \href{https://platform.openai.com/docs/assistants/tools/file-search}{file-search} interface to implement this process, where RAG is invoked by including a file search instruction in the query. The file search interface is set up using default hyper-parameter settings, such as an 800-token chunk size and a 256-dimensional embedding size.

\subsection{Planner}
Planner is a multi-modal agent responsible for identifying the next step to be executed and generate the actionable plan based on the current GUI status and the test description. 

\begin{wrapfigure}{r}{0.5\textwidth}
    \centering
    \vspace{-0.4cm}
    \setlength{\abovecaptionskip}{-0cm}
    \setlength{\belowcaptionskip}{-0.4cm}
    \includegraphics[width=0.5\textwidth]{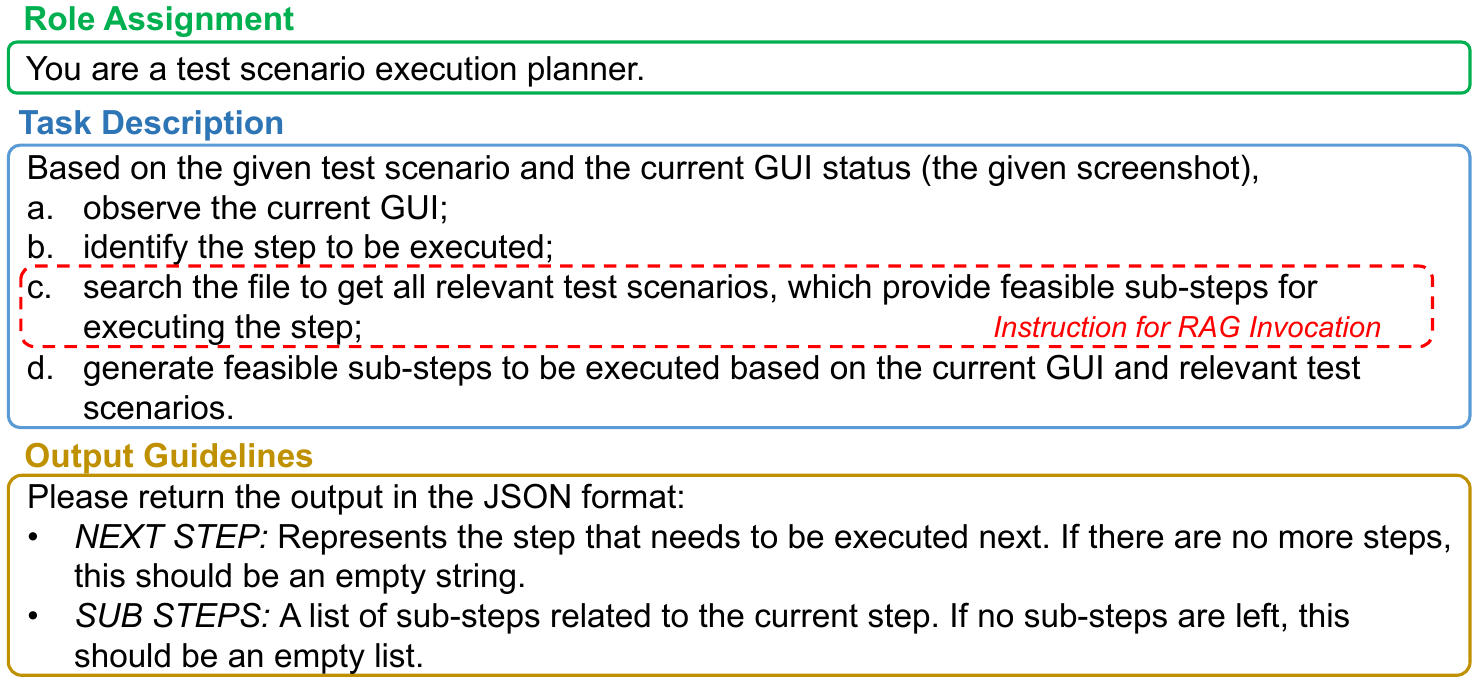}
    \caption{Role-Play Prompt for Planner Initialization}
    \label{fig:planner_prompt}
\end{wrapfigure}

\subsubsection{Agent Initialization}
As the LLMs are originally designed for general purposes, we need to specialize it for the specific role of plan generation for test steps. 
This specialization is achieved by using a system-level role-play prompt, as illustrated in Figure~\ref{fig:planner_prompt}. 
The prompt consists of three main components: \textit{Role Assignment}, which designates the LLM as a test scenario execution planner; 
\textit{Task Description}, which outlines the step-by-step instructions (four steps) to be followed, based on the chain-of-thoughts (CoTs) idea~\cite{CoTs_wei2022chain}; 
and \textit{Output Guidelines}, which directs the LLM to respond in JSON format, including fields such as ``NEXT STEP'' and ``SUB STEPS'', each with a brief description.
To incorporate crowd-sourced knowledge about the SUT when generating an actionable plan, Planner utilizes the KG-based RAG described in Section~\ref{sec:rag}. Specifically, a tailored instruction is included in the \textit{Task Description} to prompt the agent to invoke the file search interface for RAG based on the context. The retrieved step knowledge is then used to enhance the subsequent plan generation.
Additionally, Planner maintains a test-specific dialogue session that stores previous query-response records to preserve the complete testing context.



\subsubsection{Plan Generation}
Given a GUI status, Planner generates the corresponding plan based on the provided test and previous query-response records in the dialogue session. 
Specifically, a query is constructed using the template: ``Steps: [Steps]\textbf{\textbackslash n}GUI Status: [GUI Status]'', where the placeholders are filled with the given test and GUI status. After submitting the query to the agent, a response in the JSON format defined in \textit{Output Guidelines} is returned, from which sub-steps can be extracted to form the plan. 
The record of the query and response from this interaction is appended to the dialogue session for use in subsequent turns of plan generation.

\subsection{Player}
After receiving the actionable plan from Planner, Player translates the sub-step into a structured UI instruction for the current GUI status and then executes it. 

\subsubsection{Agent Initialization.}
\begin{wrapfigure}{r}{0.5\textwidth}
    \centering
    \vspace{-0.4cm}
    \setlength{\abovecaptionskip}{-0cm}
    \setlength{\belowcaptionskip}{-0.4cm}
    \includegraphics[width=0.5\textwidth]{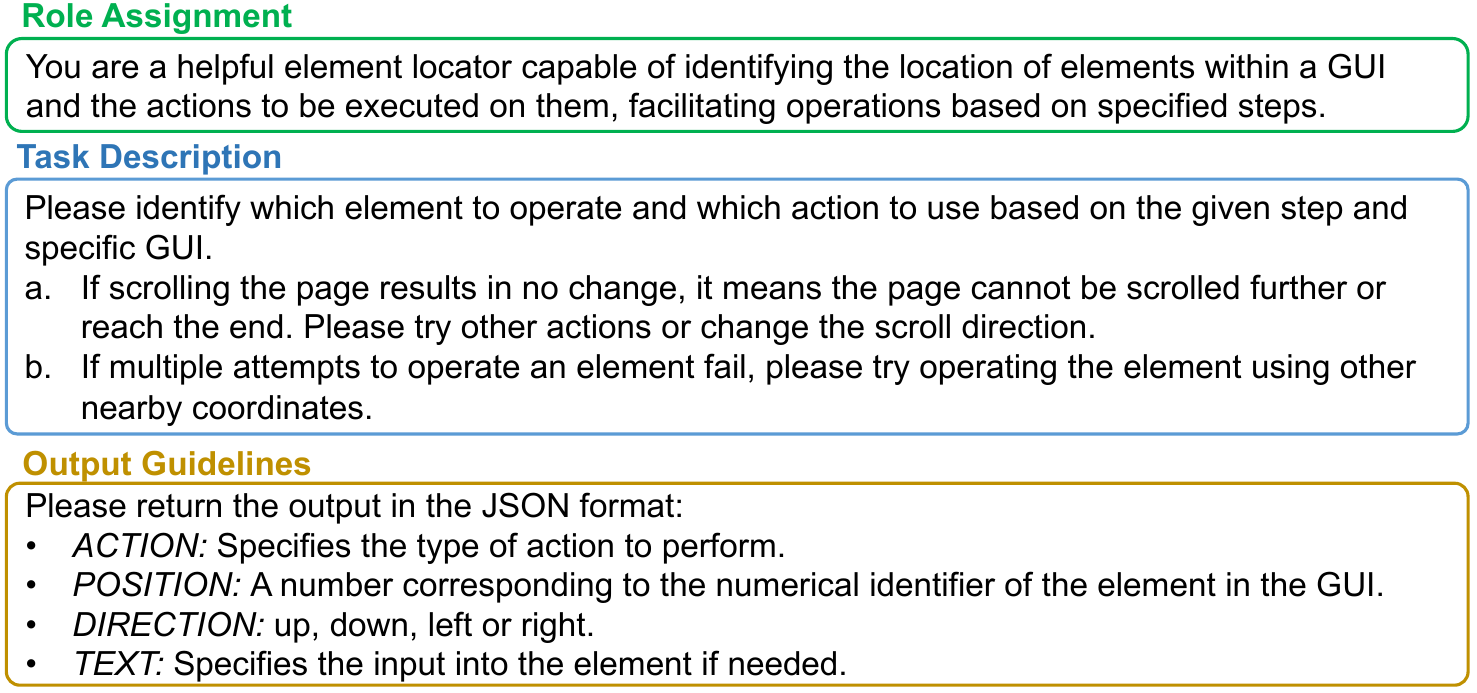}
    \caption{Role-Play Prompt for Player Initialization}
    \label{fig:player_prompt}
\end{wrapfigure}

Similar to Planner, Player is initialized with a three-component (i.e., \textit{Role Assignment}, \textit{Task Description}, and \textit{Output Guidelines}) role-play prompt, as shown in Figure~\ref{fig:player_prompt}, and maintains a dialogue session to store previous query-response records. Additionally, Player integrates a command-line tool (e.g., ADB) to execute UI instructions. We provide nine common standard actions for UI instructions: \textit{tap}, \textit{long-tap}, \textit{double-tap}, \textit{input}, \textit{scroll}, \textit{home}, \textit{enter}, \textit{landscape}, and \textit{portrait}. Specifically, \textit{home} simulates pressing the home button, \textit{enter} simulates pressing the Enter key, \textit{landscape} rotates the device to landscape mode, and \textit{portrait} switches it back to portrait mode. Less common actions, like pinch or multi-hand gestures, are excluded. Some actions require arguments: \textit{tap}, \textit{long-tap}, and \textit{double-tap} have a \textit{position} argument; \textit{input} requires \textit{position} and \textit{text}; \textit{scroll} takes a \textit{direction} argument. Other actions, such as \textit{home}, can be executed without arguments.

\subsubsection{UI Instruction Translation}
A UI instruction consists of two key components: \textit{Action}, which specifies the type of instruction, and \textit{Arguments}, which indicates the required action arguments. 


Given the plan and the current GUI status, Player constructs a LLM query using the template ``Plan: [Plan]\textbf{\textbackslash n}GUI Status: [GUI Status]'', where the placeholders are populated with the relevant information. The output is a JSON response following the format defined in the \textit{Output Guidelines}, from which a structured UI instruction is generated. The query and response from this interaction are then appended to the dialogue session for use in subsequent turns of UI instruction translation.

Before inserting the GUI screenshot into the [GUI status] placeholder in the query template, we label it with numbered elements to support actions that require a \textit{position} argument. The reasoning is as follows: A straightforward approach to represent UI positions for instructions is by directly using x-y coordinates. However, it is impractical for LLMs to predict precise widget coordinates. To address this, we divide the UI into fixed-size elements (e.g., 100px × 100px grids) and number them sequentially from left to right and top to bottom. Each element's numeric label is overlaid on the screenshot, allowing LLMs to reference the numeric labels to specify the UI position for an instruction. For example, GUI Screenshot 2 in Figure~\ref{fig:approach_workflow} shows the labeled version of GUI Screenshot 1, where the label ``32'' indicates the UI element corresponding to ``tabs tray''.

\subsubsection{UI Instruction Execution}
After generating a structured UI instruction, Player executes it using a command-line UI operation tool such as ADB~\cite{adb}. 
For actions that require a \textit{position} argument, 
the exact UI coordinates are calculated and derived from the numeric label.
As shown in Figure~\ref{fig:approach_workflow}, after executing the instruction ``[\textit{tap}] \textit{position}=32'', the GUI transitions from Screenshot 2 to Screenshot 3. 
Player captures a new screenshot to record the updated GUI status.

\subsection{Detector}
\begin{wrapfigure}{r}{0.5\textwidth}
    \centering
    \vspace{-0.4cm}
    \setlength{\abovecaptionskip}{-0cm}
    \setlength{\belowcaptionskip}{-0.4cm}
    \includegraphics[width=0.5\textwidth]{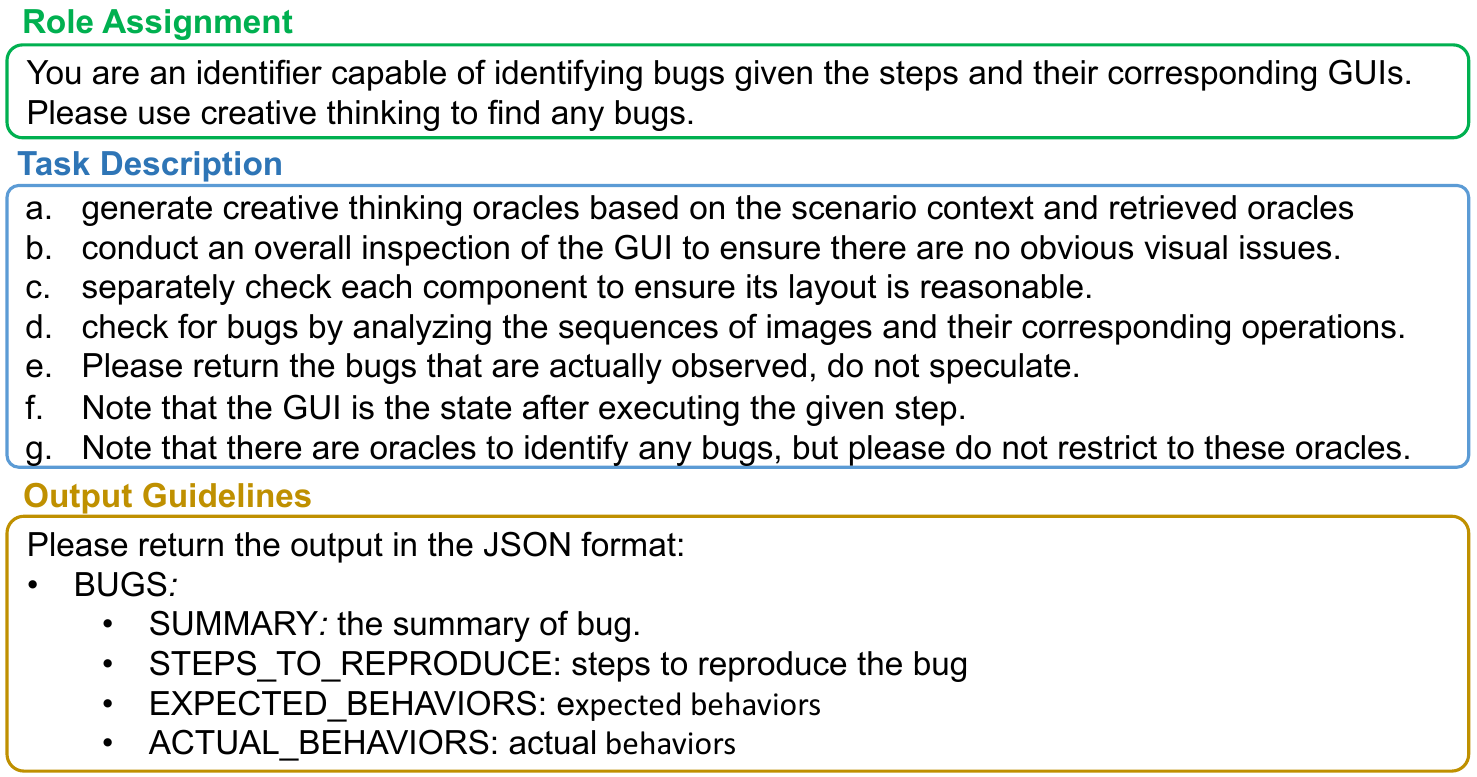}
    \caption{Role-Play Prompt for Detector Initialization}
    \label{fig:detector_prompt}
    \vspace{-0.2cm}
\end{wrapfigure}
To promptly identify unexpected behaviors, Detector retrieves relevant oracles from the SKG after each UI instruction execution and applies creative thinking to detect potential bugs.

\subsubsection{Agent Initialization}
Similar to Planner and Player, Detector is initialized with a three-component (i.e., \textit{Role Assignment}, \textit{Task Description}, and \textit{Output Guidelines}) role-play prompt, shown in Figure~\ref{fig:detector_prompt}, and maintains a dialogue session to store previous query-response records. 


\begin{wrapfigure}{r}{0.5\textwidth}
    \centering
    \setlength{\abovecaptionskip}{-0cm}
    \setlength{\belowcaptionskip}{-0.4cm}
    \includegraphics[width=0.5\textwidth]{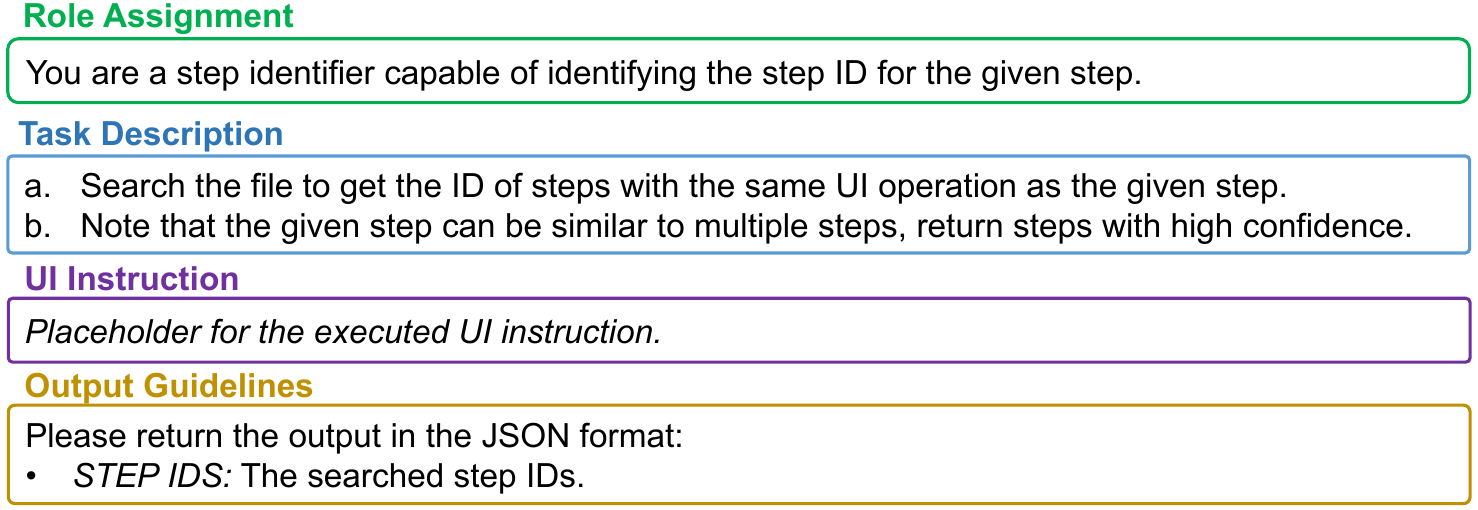}
    \caption{Prompt for Oracle Retrieval}
    \label{fig:oracle_prompt}
\end{wrapfigure}

\subsubsection{Oracle Retrieval}
We retrieve relevant oracle knowledge from SKG to assist in bug detection as follows. 
First, we extract steps similar to the executed UI instruction from the SKG using the KG-based RAG method described in Section~\ref{sec:rag}. As shown in Figure~\ref{fig:oracle_prompt}, the prompt used to invoke the RAG contains four main components: \textit{Role Assignment}, \textit{Task Description}, \textit{UI Instruction}, and \textit{Output Guidelines}. 
Using this prompt, steps similar to the executed UI instruction are returned by their unique IDs. 
Based on the retrieved steps, all associated oracles are fetched from the SKG. For example, for the UI instruction corresponding to ``Click on the tabs tray'' in Figure~\ref{fig:approach_workflow}, the retrieval returns ``Go to Tabs Tray'' and ``Open tab trays'' as similar steps, along with several associated oracles such as ``Tooltip should display explanatory text'' and ``Thumbnails should be displayed properly in the Tabs Tray''.

\subsubsection{Bug Detection}

Given the executed UI instruction and the updated GUI status, Detector performs bug detection based on the retrieved oracles and previous query-response records in the dialogue session. Specifically, a query is constructed using the template: ``UI Instruction: [UI Instruction]\textbf{\textbackslash n}GUI Status: [GUI Status]\textbf{\textbackslash n}Retrieved Oracles: [Retrieved Oracles]'', where the placeholders are filled with the given information. After submitting the query to the agent, a response is returned in the JSON format by the \textit{Output Guidelines}, from which bugs, including summary, S2Rs, EBs and OBs, can be extracted.
The record of the query and response from this interaction is appended to the dialogue session for use in subsequent turns of bug detection.

\section{Experiment}\label{sec:experiment}

To evaluate the effectiveness of automated soap opera testing and its ability to address the challenges identified in the formative study, we assessed it as follows:

\begin{description}
        \item[RQ1:] How effective is automated testing in real-time bug auto-detection?
	\item[RQ2:] How effective is automated testing in soap opera test auto-play?
	  \item[RQ3:] Is application-specific knowledge useful for automated soap opera testing?
\end{description}

\subsection{Experimental Setup}
To better understand the comparative strengths and weaknesses of automated versus manual execution,
we employed 30 soap opera tests from the formative study as our experimental dataset.

\subsubsection{Real-Time Bug Auto-Detection Setup (RQ1)}
For real-time bug auto-detection, we assess the number of bugs identified by automated testing after executing the soap opera tests and the accuracy rate of detecting valid bugs. 
Two authors independently reviewed the entire execution process, reproducing each detected bug on Android devices to verify its validity. 
Disagreements were resolved through discussion until consensus was reached. 
Both authors have extensive testing experience. 
Valid bugs were then submitted to the respective projects to collect feedback.

\subsubsection{Soap Opera Test Auto-Play Setup (RQ2)}
We evaluate how many tests automated testing can fully complete. 
Two authors independently observed the execution process to confirm whether the tests were completed. 
Since auto-play involves both Planner and Player, we conducted a deeper analysis of incomplete executions. 
Out of 30 soap opera tests, 325 Planner-Player executions occurred. 
We randomly sampled 50 pairs for evaluation.
Planner was assessed on accuracy of identifying the next step and 
performance of sub-steps generation using a Likert scale~\cite{likert} where 1 is the worst and 5 is the best.
For Player, we evaluated accuracy of parsing steps into structured UI instructions and accuracy of element location.
Discrepancies were resolved through discussion.

\subsubsection{Application-Specific Knowledge on Automated Soap Opera Testing Setup (RQ3)}
We conducted an ablation study to investigate the impact of application-specific knowledge (steps and oracles in SKG) on both the auto-play and bug detection for soap opera testing. 
For step knowledge used in Planner,
we disabled the step retrieval feature and examined the resulting execution outcomes, along with the number and validity ratio of detected bugs.
For oracle knowledge used in Detector, to ensure the fairness, we reused existing results from Planner and Player, rerunning only the Detector with its oracle retrieval disabled. We then analyzed the new numbers of detected bugs.





\subsection{Effectiveness in Real-Time Bug Auto-Detection (RQ1)}

\begin{wraptable}{r}{0.43\textwidth}  
\caption{Comparison of Detected Bug Numbers}
\label{tab:bug_finding}
\centering
\renewcommand{\arraystretch}{1.1}
\resizebox{0.43\textwidth}{!}{  
\begin{tabular}{l|c|c|c}
\toprule
\textbf{Approach}     & \textbf{TP \#} & \textbf{FP \#} & \textbf{Acc.} \\ 
\midrule
Firefox (Manual$^1$)      & 27             & 0              & 1.000 \\
Firefox (Automated)       & 32             & 30             & 0.516 \\
\hdashline 
WordPress (Manual$^1$)    & 14             & 0              & 1.000 \\
WordPress (Automated)     & 21             & 11             & 0.656 \\
\hdashline 
AntennaPod (Manual$^1$)   & 16             & 0              & 1.000 \\
AntennaPod (Automated)    & 15             & 11             & 0.577 \\
\bottomrule
\multicolumn{4}{l}{\footnotesize $^1$Manual Soap Opera Testing from Formative Study}
\end{tabular}
}
\end{wraptable}

As illustrated in Table~\ref{tab:bug_finding}, automated testing has shown promising capabilities in real-time bug auto-detection. 
32, 21 and 15 bugs were confirmed by the two authors for Firefox, WordPress and AntennaPod respectively.
Due to time and resource limitations, we selected 15, 9, and 10 bugs for submission to their respective projects, prioritizing defects over enhancements. 
For enhancements, we focused on those that significantly impacted user experience. 
Each report was meticulously crafted with detailed descriptions and accompanied by screenshots or videos for reference, ensuring quality over quantity.
After reporting
the 15+9+10
bugs, 
3 bugs have been fixed, 13 confirmed, 6 assigned priority and severity, 1 marked as duplicates.
For example, automated testing identified an unexpected \href{https://github.com/AntennaPod/AntennaPod/issues/7365}{Issue7365} (marking the episode as unplayed in `Manage downloads' does not update the contrast to its unplayed status) during the execution, which has been fixed.
Simultaneously, automated testing has also exposed areas that urgently need improvement: enhancing the capability to exploration boundary and reducing the rate of incorrectly identified bugs (false positives).

\subsubsection{Under-Explored Scenario Boundary}
Looking solely at the number of true positive bugs detected in Table~\ref{tab:bug_finding},
automated testing appears to yield comparable results to manual soap opera testing conducted by participants.
However, by delving deeper into the bugs detected, we discovered that automated testing and manual testing tend to identify different bugs. 
Among all the bugs identified by both automated testing and manual testing, only two were identical. 
This shows the effectiveness of soap opera testing in finding diverse bugs, but also reveal the varying capabilities of manual testing versus automated testing.
Automated testing demonstrates greater sensitivity to enhancements, focusing on aspects such as user usability and design consistency, whereas participants tend to focus more on detecting defects, such as functionality, crashes.
Notably, a significant portion of the true-positive bugs identified by automated testing were enhancements: 19 out of 32 for Firefox, 15 out of 21 for WordPress, and 11 out of 15 for AntennaPod. 
In comparison, participants in the formative study identified fewer enhancement bugs: 3 out of 27 for Firefox, 1 out of 14 for WordPress, and 2 out of 16 for AntennaPod.
An in-depth comparison of automated and manual soap opera testing revealed that automated testing requires improvement in expanding boundary exploration through creative thinking.
This is crucial, as some defects can only be discovered by exploring beyond the initial test scenarios.

As detailed in Section~\ref{subsubsec:insights2_expand_boundaries}, during the initial soap opera test derived from \href{https://bugzilla.mozilla.org/show_bug.cgi?id=1816146}{Bug1816146} on editing a saved login, $P1_{Firefox}$ identified two defect bugs by expanding the boundaries of exploration.
Upon comparison, automated testing generates mostly conventional inputs, using typical examples such as usernames and passwords: `your\_username' and `password'. 
This input generation might not effectively simulate real-world user behaviors or thoroughly test the boundaries of the system's error handling capabilities.  
Thus, automated testing highlighted several enhancements during this testing. 
On `Logins and passwords' page, automated testing identified an inconsistency where the `Exceptions' button features an icon, diverging from other sections that use text only (\href{https://bugzilla.mozilla.org/show_bug.cgi?id=1912617}{Bug1912617}).
On `Saved passwords' page, it suggested that the full text of saved password entries should be either visible or accessible through text wrapping or a tooltip, enhancing accessibility (\href{https://bugzilla.mozilla.org/show_bug.cgi?id=1912621}{Bug1912621}).
On `Add password' page, automated testing discovered that the URL validation message remains static, failing to update and provide feedback based on the user's input
 (\href{https://bugzilla.mozilla.org/show_bug.cgi?id=1907851}{Bug1907851}). 
 Additionally, the `Password' field on the same page lacks a `Show/Hide' toggle, a feature expected for user convenience (\href{https://bugzilla.mozilla.org/show_bug.cgi?id=1912202}{Bug1912202}).
 After reporting these bugs, one was confirmed, and two were assigned priority or severity levels.
 Inspired by \href{https://bugzilla.mozilla.org/show_bug.cgi?id=1912202}{Bug1912202}, an author found that `Site' field misses `Clear' button for removing invalid websites
 (\href{https://bugzilla.mozilla.org/show_bug.cgi?id=1912200}{Bug1912200}).
Input exploration can be enhanced by developing an input generation module that leverages the GUI state and applies testing techniques like boundary value analysis to generate diverse inputs, thereby expanding the input exploration boundaries.


\subsubsection{Incorrectly Identified Bugs}

From Table~\ref{tab:bug_finding}, it is evident that automated testing exhibits a relatively high false positive rate in bug detection. 
To understand the underlying reasons for this high rate, we conducted a detailed analysis of all 52 incorrectly reported bugs and identified four main causes: LLMs hallucination, screenshot-based GUI presentation, error GUI element location, and unreasonable or unnecessary enhancement suggestions.

\textbf{LLMs Hallucination (14 out of 52: 26.9\%).}
Of the 52 incorrectly reported bugs, 14 were attributed to hallucinations from LLMs, which break down into two categories: hallucinations about UI design issues (9/14) and unexecuted execution plans (5/14).
These hallucinations can arise either directly from the LLMs themselves or from the application of oracle knowledge.

\textit{False UI Design Bugs (9 out of 14):} Hallucinations in this category include issues such as overlap, obscuring, and GUI design inconsistency. 
For instance, the LLMs might creatively generate an oracle like ``Check if the save icon is enabled even when required fields are empty'', and then erroneously hallucinates a non-existent bug: the save icon is reported as enabled while, in fact, it is disabled. 
Hallucinations can also be prompted by oracle knowledge from the scenario knowledge base. 
For example, when the user taps on the three-dot menu in the top right corner, an oracle is retrieved for bug detection stating, ``Reduce the height of the 3-dot menu by consolidating some items to make it fit within the screen without requiring scrolling''.
Influenced by this oracle, the LLM mistakenly perceives that ``The menu appears to extend beyond the screen's height, which might result in a need for scrolling, especially on smaller screens'', even though this issue does not actually exist.

\textit{Unexecuted Execution Plans (5 out of 14):} 
These hallucinations occur when the LLM predicts bugs from actions or sequences that were never executed during testing. 
Hallucinations influenced by oracle knowledge are particularly intriguing. 
For instance, an oracle in WordPress states, ``The `No internet' snackbar should appear when the bottom loader view becomes visible''. 
Due to this mention of ``no internet'', Detector mistakenly generates a bug report without executing the last two steps described in the S2Rs: ``Open Stats section'', ``Scroll to map view with network'', ``Switch to offline mode'', and ``Observe response and rendering''. 
The mention of ``no internet'' prompted an unexecuted operation related to switching to offline mode, causing a false positive.
However, this oracle can be used strategically to guide LLMs in generating new test branches, expanding the testing scope. 
By executing steps generated from this oracle, we can assess whether the current scenario reveals issues when the system operates offline. 
This highlights how oracle knowledge can be leveraged to create new branches, allowing for more comprehensive testing of system behavior.
This is particularly effective with oracles that include actionable steps, such as ``The vertical search menu/context menu should be dismissed by tapping outside or using the back button''. 


\textbf{Screenshot-based GUI Presentation (12 out of 52: 23.1\%).}
Misreporting can be further divided into two situations: the first occurs due to incomplete rendering/loading at the time of the screenshot, while the second arises from a failure to capture dynamic changes.
For instance, a false positive bug from WordPress reported, ``Post creation screen does not display any text input fields or content creation tools'', due to incomplete rendering/loading.
To address this issue, we can invoke adaptive screenshoting tools like AdaT~\cite{fengICSE2023GUIrenderingInference}, which dynamically adjusts the screenshoting timing based on the GUI rendering state to
reduce false positives.
An example of a false positive caused by failure to capture dynamic changes is ``Episodes list does not update after tapping `Refresh' ''. 
In reality, the loading animation appeared after tapping ``Refresh'', but was not visible in the screenshot due to the screenshoting timing.
To address this, 
video-based bug detection approach or mitigating the issue by taking more frequent and dense screenshots can be adopted. 

\textbf{Error GUI Element Location (6 out of 52: 11.5\%).}
Another type of false bug report is caused by GUI element location errors.
For instance, the report ``Security information panel does not display after tapping the lock icon'' was actually due to the lock icon's incorrect coordinates, which caused the icon to not be clicked properly. 
To address this, we need to improve the accuracy of element positioning, potentially by fine-tuning the LLMs for better spatial understanding and interaction.

\textbf{Unreasonable or Unnecessary Improvement Suggestions (20 out of 52: 38.5\%).} 
This type of false bug report arises from generating improvement suggestions that are either unreasonable or unnecessary for the system's actual functionality. 
Such suggestions often contribute little or are based on misunderstandings.
For example, the suggestion, ``The `Sponsored' label for Amazon and Hotels.com is inconsistent with other site labels'', is unreasonable since only these two websites are sponsors, making the call for additional labels unreasonable. 
Similarly, the suggestion ``Lack of sorting or filtering options for the subscriptions page'' was unnecessary, as these options are already accessible via the three-dot icon on the page. 
These suggestions generally result from a lack of understanding of the app's layout and business context.
To address this, we plan to develop a multimodal KG that integrates GUI information as a reference. Additionally, the KG should incorporate a broader range of information sources, such as requirement manuals, design documentation, and other business-relevant resources
to provide more comprehensive context.

\subsection{Effectiveness in Soap Opera Test Auto-Play (RQ2)}

We found that 7 out of 10 soap opera tests were fully executed for Firefox, and 6 out of 10 for both WordPress and AntennaPod. 
Planner achieved 92.0\% accuracy in identifying the next step, with sub-step generation receiving positive feedback, 72.0\% rated `5' (best) on a Likert scale, and only 10.0\% rated `2' or `1' (worst). 
Player had an 84.0\% accuracy in parsing steps into structured UI instructions and a 70.0\% accuracy in element location. 
While the Planner performed well, improvements are needed in the Player’s element location accuracy to enhance overall test execution.

In the evaluation, we found that Planner demonstrated the ability to dynamically adjust execution plans in real time, based on the specific state of the GUI. 
This adaptability allows the Planner to respond effectively to changes in the interface, enhancing its performance in guiding the soap opera test auto-play and even expanding the boundaries of exploration.
As shown in Figure~\ref{fig:approach_workflow}, after closing a tab, the test requires reopening a closed tab. During execution, the GUI briefly displays a ``Tab closed UNDO'' banner
after the tab is closed. 
Planner adapts by deciding to use the `UNDO' feature after detecting the tab closure, making it easier to reopen the recently closed tab. 
However, since the banner only appeared for a short time, by the time the Player attempted to execute the action, the banner had vanished, preventing the operation from completing.
After the result analysis, observing this branch exploration inspired the author to discover two new bugs:
\href{https://bugzilla.mozilla.org/show_bug.cgi?id=1912742}{Bug1912742} (The recently closed tab reappears in its previous position but without the content preview when using the `UNDO' button in the tabs tray) and \href{https://bugzilla.mozilla.org/show_bug.cgi?id=1912747}{Bug1912747} (The `UNDO' action highlights the last two tabs instead of just the last one on the tabs tray). 
Both bugs have been confirmed.
This unveils the opportunity for human-AI collaboration for soap opera testing.

Although the Player demonstrated suboptimal performance in element location accuracy, we unexpectedly found that the inaccurate positioning contributed to the exploration of random branches. 
For instance, while executing the last step in the soap opera test shown in Figure~\ref{fig:approach_workflow}, after selecting a tab, the Player intended to click the Three-dot menu but mistakenly positioned the click on the Share button.
This opens the share page and unintentionally explores a random branch.
Thanks to the adaptive capabilities of the Planner, it recognized that the Share operation was unrelated to the current test.
Then, Planner generated a plan to close the Share page, effectively returning to the main task. 
However, this detour into the random branch led to the discovery of an unexpected bug: ``The back icon turns black and the title reverts to `Recently Closed Tabs' while still displaying the selected websites after canceling a share action on the `Recently Closed Tabs' page''.
This bug was identified due to the random branch exploration and was reported and confirmed as \href{https://bugzilla.mozilla.org/show_bug.cgi?id=1912905}{Bug1912905}.
After analyzing the bug discovery process, inspired by this branch exploration, the author applied equivalence class to test similar branches in other pages with selection and share functionality.
As a result, two similar bugs were found on the Tabs Tray page and Bookmarks page, reported as \href{https://bugzilla.mozilla.org/show_bug.cgi?id=1912910}{Bug1912910} (confirmed) and \href{https://bugzilla.mozilla.org/show_bug.cgi?id=1912912}{Bug1912912} (assigned priority and severity).

This example highlights two key points. 
First, Planner can ensure that even when exploring random branches, it can still return to and complete the main task. 
Second, Player's inaccurate element positioning shows how unintentional actions can lead to the discovery of new issues.
In future iterations of automated soap opera testing, introducing a certain degree of randomness could help expand exploration boundaries and trigger unexpected bugs.
Besides,
inspired by the exploratory execution of automated testing, the authors identified 16 additional bugs, where 1 crash was fixed, 8 confirmed, 2 assigned priority and severity, 1 marked as a duplicate, and 1 as wontfix.



\subsection{Usefulness of Application-Specific Knowledge (RQ3)}
\begin{wraptable}{r}{0.5\textwidth}  
\vspace{-0.4cm}
\caption{Ablation Study on Bug Detection Accuracy}
\setlength{\abovecaptionskip}{0.2pt}  
\setlength{\belowcaptionskip}{-0.2pt}  
\label{tab:ablation}
\vspace{-2mm}
\centering
\renewcommand{\arraystretch}{1.1}
\resizebox{0.5\textwidth}{!}{  
\begin{tabular}{l|c|c|c|c|c|c|c|c|c}
\toprule
\multirow{2}{*}{\textbf{Tool}}                               & \multicolumn{3}{c|}{\textbf{Firefox}}                                                      & \multicolumn{3}{c|}{\textbf{WordPress}}                                                    & \multicolumn{3}{c}{\textbf{AntennaPod}}                                                    \\ 
\cline{2-10}
                                  & \multicolumn{1}{c|}{TP \#} & \multicolumn{1}{c|}{FP \#} & Acc. & \multicolumn{1}{c|}{TP \#} & \multicolumn{1}{c|}{FP \#} & Acc. & \multicolumn{1}{c|}{TP \#} & \multicolumn{1}{c|}{FP \#} & Acc. \\ 
\midrule
Automated Testing                      & 32           & 30            & 0.516    & 21           & 11            & 0.656    & 15           & 11            & 0.577    \\ 
- {\scriptsize\textit{w/o} Oracle Knowledge} & 8            & 29            & 0.216    & 8            & 11            & 0.421    & 9            & 19            & 0.321    \\ 
- {\scriptsize\textit{w/o} Step Knowledge}   & 10           & 23            & 0.303    & 7            & 35            & 0.167    & 13           & 26            & 0.333    \\ 
\bottomrule
\end{tabular}
}
\vspace{-2mm}
\end{wraptable}
As shown in Table~\ref{tab:ablation}, without the support of oracle knowledge, the number of effectively identified true-positive bugs significantly decreases, and the proportion of true bugs drops noticeably, specifically, from 0.516 to 0.216 for Firefox, 0.656 to 0.421 for WordPress, and 0.577 to 0.321 for AntennaPod. 
This highlights the critical role that oracle knowledge plays in enhancing real-time bug auto-detection.
Even so, the oracle knowledge from SKG still has room for improvement.  
First, oracles with similar meanings often appear repeatedly. 
For example, as shown in Figure~\ref{fig:approach_workflow}, after executing "Click on the tabs tray," the retrieved oracles include "Tooltip should display explanatory text" and "View the tooltip with missing text," which are essentially duplicates. 
These could be merged using semantic clustering to avoid redundancy.
Second, some oracles are irrelevant to the current testing scenario. 
In the same example, one of the retrieved oracles is "The thumbnail image for the image opened in a new tab should be correctly displayed," which is irrelevant because no new tab has been opened in the current soap opera test. 
However, these oracles are beneficial for generating new branch to expand the exploratory boundary.
Lastly, for infeasible oracles, such as "Locate the tab switcher icon inside the tab tray" (where the current GUI no longer uses this design), filtering mechanisms should be implemented to remove these infeasible oracles, which could help reduce hallucinations and improve the accuracy of bug detection.

Without the support of step knowledge, the number of fully executed soap opera tests dropped significantly, from 7 to 3 for Firefox, from 6 to 3 for WordPress, and from 6 to 5 for AntennaPod. 
As shown in Table~\ref{tab:ablation}, 
this also significantly reduced the effectiveness of real-time bug auto-detection, with detection rates dropping from 0.516 to 0.303 for Firefox, 0.656 to 0.167 for WordPress, and 0.577 to 0.333 for AntennaPod.
In the analysis of step knowledge, we found that the granularity and breadth of the steps have a significant impact on the quality of automated soap opera testing.
For example, in Firefox's knowledge base, the scenario sources are extensive and come from a large user base, including developers, testers, and regular users, all of whom have varying writing habits. 
The advantage of this diversity is that while some scenarios are written with poor quality, often missing UI instructions, other contributors' writings help fill in these gaps. 
However, for an app like AntennaPod, which has fewer users writing issues and pull requests, as well as a smaller development team, the missing UI instructions are not as readily supplemented.
In contrast, for WordPress, the scenarios often come from pull requests containing high-quality, fine-grained steps written by developers. 
This is beneficial for auto-play, as it provides clear, detailed instructions, but the downside is that some features are not covered, making it inaccessible and limiting the effective exploration of certain features.
For example, as seen in Table~\ref{tab:ablation}, the omission of step knowledge in WordPress led to the most significant drop in TP and a substantial increase in FP.
To address these issues and further tackle the cold-start problem, a UI auto-exploration model could be designed to automatically construct a comprehensive KG of the app's functionality and UI operations.

\section{Road Ahead for Automated Soap Opera Testing}\label{sec:insights}
Based on the aforementioned research, we found that automated soap opera testing directed by LLMs and scenario knowledge shows promising performance. 
However, several issues have also emerged that need to be addressed. 
In this section, we summarize the observations and outline the road ahead for automated soap opera testing.
\subsection{The Synergy of Neural and Symbolic Approaches}
Based on the experimental results, effective soap opera testing with LLMs requires comprehensive application-specific knowledge. 
Therefore, combining neural and symbolic approaches is essential.

Even though crowdsourced information (e.g., bug reports, user manuals) provides a valuable resource, it often faces limitations in scale and detail, resulting in challenges such as: a) abstract step descriptions, making direct execution difficult;
b) limited knowledge, leading to incomplete coverage of certain software functionalities.
This issue is particularly significant in newer software projects that lack sufficient crowdsourced contributions. 
As a solution, generation of functionality KG of UI instructions by automated software exploration~\cite{software_exploration_wang2023enabling,software_exploration_wen2023empowering,software_exploration_zhang2023responsible} offers a promising solution.

When LLMs perform soap opera testing, their ability to expand the exploration boundary through creative thinking is still basic,
especially when combined with software engineering testing techniques. 
This limitation leads to many bugs not being triggered. 
Therefore, how to assist LLMs in generating exploration boundaries using diverse knowledge is a direction that requires further exploration.
Exploration boundaries can be divided into two types: input exploration and branch exploration.
For input exploration, we can structure the input field as a separate node when constructing the KG. 
The KG can integrate previously existing input values from various crowdsourced sources, such as bug reports.
Further, by combining testing techniques with the LLMs' GUI understanding capabilities, various representative inputs can be generated based on the input context. 
This approach enables effective input boundary exploration, ensuring a comprehensive evaluation of possible input scenarios.
For branch exploration, we can enhance the KG by abstracting scenarios and integrating similar ones at a higher level, allowing LLMs to perform equivalence class reasoning and logically explore related branches. 
Additionally, oracles can be used to guide LLMs in generating effective branches. 
Random branch exploration (e.g., accidentally tap the wrong button)
may also yield unexpected results, so during soap opera test execution, we can introduce random factors via LLMs to facilitate random exploration.
By utilizing GraphRAG, we can better leverage the structured information within the KG to enhance this process.
It is important to ensure that boundary exploration is conducted while completing the main tasks, to avoid getting lost in creative exploration. However, determining the most effective balance between creativity and precision is still an unresolved issue. 
Therefore, exploring the creativity and precision trade-off and identifying the optimal balance is a very important and interesting research direction.

By capturing and analyzing video streams of GUI interactions, we can identify bugs missed by static snapshots or logs, particularly issues with transitions, animations, and dynamic elements. 
A key challenge is processing large volumes of video data while synchronizing it with UI instructions. 
Overlaying UI instructions as subtitles in the video can improve clarity and help identify anomalies.

\subsection{Human-AI Co-Learning}
We found that not only can humans provide insights for AI in automated soap opera testing, but AI's behavior can also inspire human creativity, leading to the discovery of more bugs. 
Inspired by automated soap opera testing, the authors identified 16 additional bugs, where 1 crash was fixed, 8 were confirmed, 2 were assigned priority and severity, 1 was marked as a duplicate, and 1 as wontfix.
Additionally, we observe some false positive situations caused by LLM hallucinations are a challenge for LLMs, but humans can quickly distinguish between true and false bugs.
For instance, a false bug such as ``The `Done' button appears slightly larger and misaligned...'' is easily identifiable as incorrect by a human tester. 
Given the above two findings, as we transition from manual to automated soap opera testing, we could adopt a middle-ground approach by designing a Human-AI collaborative intelligence method
to assist the process. 
On one hand, this can reduce the incorrectly identified bugs 
caused by LLM hallucinations, and on the other hand, human involvement may also be inspired by the automated execution process, leading to the discovery of even more bugs.
The challenge with this approach lies in when and how humans should intervene and in which way human and AI can effectively collaborate.
For example, we can combine program analysis, testing and LLM-based inference
to generate a confidence score for identified bugs, and when the score falls below a certain threshold, human intervention would be triggered.

\subsection{Soap Opera Testing Integrated with Broader Software Engineering Knowledge}
Instead of stopping after discovering a single bug, soap opera testing excels at continuing to explore around the discovered bug, uncovering various related issues, and even identifying different manifestations of the same root cause. 
For example, in \href{https://github.com/AntennaPod/AntennaPod/issues/7349}{Issue7349} and \href{https://github.com/AntennaPod/AntennaPod/issues/7350}{Issue7350}, one involved infinite loading, and the other a crash. 
After reporting both, the developer discovered they stemmed from the same underlying issue.
To improve this process, we propose integrating soap opera testing with broader software engineering knowledge, such as code, by constructing a scenario-code linked KG. 
This KG would link scenarios to code, helping identify shared root causes when multiple bugs point to the same code segment. 
This approach enables concentrated bug fixing, addressing root causes instead of individual bugs.
Additionally, we can leverage the different scenarios linked to the same code segment to recommend branches for further exploration.
The challenge in this research direction lies in how to accurately and finely restore the traceability between scenarios and code.







\section{Threats to validity}\label{sec:threats}
To mitigate potential human bias due to the small number of participants in the formative study, we recruited professionals with at least two years of experience in soap opera testing of large interactive systems with GUIs, such as desktop or mobile applications.
To further reduce bias during the evaluation, two authors independently assessed the results, discussing disagreements to reach consensus.
In cases of disagreement, they discussed the differences to reach a consensus. 
Additionally, we reported the bugs identified in both the formative study and automated soap opera testing to the respective open-source repositories for developer confirmation.

A significant external validity threat is that the evaluation was conducted on only three applications.
However, these applications
span different categories. 
The diversity of their functionalities helps mitigate potential bias.
We plan to deploy automated soap opera testing on a broader range of software systems to further validate its feasibility. 
The KG construction relies on crowdsourced information. 
For newer or less popular software that lacks such data, we aim to develop an automated software exploration method to generate a functionality KG of UI instructions to guide automated soap opera testing.

\section{Related Work}\label{sec:related}
Manual exploratory testing has been shown to be effective at the system testing level~\cite{test_computer_software_by_kanerCem,exploratory_nature,exploratory_tester_knowledge, vaaga2002managing, lyndsay2003adventures_session_based_testing,wood2003applying,bach2000session_based_test_management,bach2004exploratory,An_experiment_on_the_effectiveness_efficiency_of_ET}. 
Principles and guidelines for conducting and managing exploratory testing have been proposed~\cite{bach2000session_based_test_management, exploratory_book_tips_tricks_tours_techniques_to_guide_test_design, cem2013introduction_to_scenario_testing,soap_opera_testing, lyndsay2003adventures_session_based_testing}.
Scenario-based exploratory testing, also known as soap opera testing~\cite{soap_opera_testing}, involves designing complex scenarios to provoke failures. 
Recent work~\cite{syskg, soapoperatg, su2024enhancing} has supported soap opera testing by constructing a system KG (SYSKG) of user tasks and failures and generating soap opera tests by combining relevant scenarios based on the SYSKG. 
SoapOperaTG~\cite{soapoperatg} is developed to optimize the usage of soap opera tests.
Experimental results~\cite{syskg, soapoperatg, su2024enhancing} show that these generated soap opera tests are highly effective in identifying bugs. 
However, the manual execution required has hindered the large-scale adoption.
Thus, our work investigates the feasibility, challendges and road ahead for automated soap opera testing directed by LLMs and scenario knowledge.

Given the remarkable performance of LLMs in NLP and vision, researchers have begun to explore their applicability or impact on software engineering, such as automated code completion~\cite{llmsJain2022jigsaw,llmsNashid2023retrieval}, software development~\cite{llmsImai2022github,llmsSarkar2022like}, security vulnerabilities~\cite{llmsGrishina2022enabling,LLMVulpearce2023examining}, software exploration~\cite{software_exploration_wang2023enabling,software_exploration_wen2023empowering,software_exploration_zhang2023responsible} and software testing~\cite{liu2023_input_generation, liu2024testing_unsual_text_inputs,axnav_taeb2024}.
However, no work has yet explored the use of LLMs to automate soap opera testing.
Maryam et al.~\cite{axnav_taeb2024} developed AXNav, a multi-agent system leveraging multi-modal LLMs to interpret accessibility tests in NL and replay the instructions on mobile devices while manipulating relevant accessibility features. 
AXNav currently detects issues related to four accessibility features: VoiceOver, Dynamic Type, Bold Text, and Button Shapes.
Unlike AXNav, which focuses on replaying accessibility tests for specific accessibility issues, our work explores automated soap opera testing, which is not confined to specific bug types. 
It enables the detection of any bugs, even those beyond predefined oracles, during the execution of soap opera tests. 
Furthermore, it allows to vary the tests as appropriate, enhancing both the breadth and depth of the testing process.

Recent work has investigated reproducing bugs from the S2Rs in bug reports and issues~\cite{automatically_translate_bug_into_test_case,ReCDroid_zhao2019recdroid,sidong2024prompting} or from bug videos~\cite{bernal2020translating_video_recordings,havranek2021_v2s_translating_video_recordings_tool}. 
Su et al.~\cite{soap_opera_testing,soapoperatg,syskg} have extracted useful sections, such as S2Rs and EBs, to generate soap opera tests. 
BUGINE~\cite{collaborative_bug_finding,li2020bugine_collaborative_bug_finding} recommends bugs from certain apps to developers of similar apps for testing. 
None of these approaches leverage the abundant scenario knowledge present in bug reports, issues, and pull requests to guide automated soap opera testing.



\section{Conclusions}\label{sec:conclusion}
This paper identifies key insights for effective manual soap opera testing and the challenges in automating this process through a formative study. 
Based on these insights and challenges, we develop a multi-agent system that leverages LLMs and a scenario knowledge graph to automate soap opera testing. 
The system comprises three multi-modal agents, Planner, Player, and Detector, which collaborate to execute soap opera tests and identify potential bugs. 
Experimental results demonstrate the feasibility of automated soap opera testing, with identified bugs being confirmed and fixed by the development team. 
However, there remains a gap in achieving the effectiveness of manual soap opera testing, particularly in addressing under-explored scenario boundaries and reducing incorrectly identified bugs (false positives). 
We further discuss opportunities, challenges and potential solutions in the road ahead for automated soap opera testing, focusing on three key areas: the synergy of neural and symbolic approaches, human-AI co-learning, and integrating soap opera testing with broader software engineering practices.

\section*{Data Availability}\label{sec:data}
Replication package with code and dataset: \url{https://github.com/SoapOperaTester/SoapOperaTesting}

\bibliographystyle{ACM-Reference-Format}
\bibliography{acmart}










\end{document}